\documentclass[journal]{IEEEtran}
\usepackage[cmex10]{amsmath}
\usepackage{color}
\usepackage{longtable}
\usepackage{xcolor}
\usepackage{blindtext}
\usepackage{upgreek}

\usepackage{amssymb,bezier,amsfonts, enumerate, keyval}
\usepackage[cmex10]{amsmath}
\usepackage{amssymb,bezier,times} 
\usepackage{color}
\usepackage[shortlabels]{enumitem}
\usepackage{multirow}
\usepackage{float}
\usepackage[caption = false]{subfig}
\usepackage[final]{graphicx}
\usepackage{amsthm}

\usepackage[ruled,linesnumbered,norelsize]{algorithm2e}
\makeatletter
\newcommand{\removelatexerror}{\let\@latex@error\@gobble}
\makeatother

\theoremstyle{definition}

\makeatletter
\let\oldlt\longtable
\let\endoldlt\endlongtable
\def\longtable{\@ifnextchar[\longtable@i \longtable@ii}
\def\longtable@i[#1]{\begin{figure}[t]
\onecolumn
\begin{minipage}{0.5\textwidth}
\oldlt[#1]
}
\def\longtable@ii{\begin{figure}[t]
\onecolumn
\begin{minipage}{0.5\textwidth}
\oldlt
}
\def\endlongtable{\endoldlt
\end{minipage}
\twocolumn
\end{figure}}
\makeatother

\ifCLASSINFOpdf
\else
\fi
\begin{document}
%
\title{Scalable Multi-Party Privacy-Preserving Gradient Tree Boosting over Vertically Partitioned Dataset with Outsourced Computations}
%
%
%

\author{Kennedy~Edemacu,
        Beakcheol~Jang,
        Jong~Wook~Kim
\thanks{Kennedy Edemacu is with the Department of Computer Science and Electrical Engineering, Muni University, Arua, Uganda}
\thanks{Beackcheol Jang is with the Graduate School of Information, Yonsei University, Seoul, South Korea}
\thanks{Jong Wook Kim is with the Department of Computer Science, Sangmyung University, Seoul, South Korea}
\thanks{E-mail~addresses: k.edemacu@muni.ac.ug, bjang@yonsei.ac.kr, jkim@smu.ac.kr}
\thanks{Jong Wook Kim is the corresponding authors.}}
%
%
%

\markboth{}%
{Edemacu \MakeLowercase{\textit{et al.}}: Scalable Multi-Party Privacy-Preserving Gradient Tree Boosting over Vertically Partitioned Dataset with Outsourced Computations}
%



\maketitle

\begin{abstract}
Due to privacy concerns, multi-party gradient tree boosting algorithms have become widely popular amongst machine learning researchers and practitioners. However, limited existing works have focused on vertically partitioned datasets, and the few existing works are either not scalable or tend to leak information. Thus, in this work, we propose SSXGB which is a scalable and secure multi-party gradient tree boosting framework for vertically partitioned datasets with partially outsourced computations. Specifically, we employ an additive homomorphic encryption (HE) scheme for security. We design two sub-protocols based on the HE scheme to perform non-linear operations associated with gradient tree boosting algorithms. Next, we propose a secure training and a secure prediction algorithms under the SSXGB framework. Then we provide theoretical security and communication analysis for the proposed framework. Finally, we evaluate the performance of the framework with experiments using two real-world datasets.      
\end{abstract}

\begin{IEEEkeywords}
Gradient tree boosting, multi-party machine learning, privacy-preservation, homomorphic encryption, vertically partitioned dataset. 
\end{IEEEkeywords}

%
\IEEEpeerreviewmaketitle

\section{Introduction}\label{sec:introduction}
%
%
%
%
\IEEEPARstart{T}{he} privacy-preserving multi-party machine learning paradigm has shown promising potential in encouraging collaboration between organizations while preserving the privacy of their data \cite{gong2020}. The basic idea of the privacy-preserving multi-party machine learning is that each collaborating party holds a private dataset and trains a local model using the dataset. The local models from the participating parties are then aggregated to create a single more powerful model. Hence, different organizations can jointly train a machine learning model without sharing their private datasets. 
   
Although the privacy-preserving multi-party machine has attracted a lot of attention recently, the majority of the existing works focus on linear regression \cite{cock2015, hall2011}, logistic regression \cite{kim2018, aono2016} and neural networks \cite{zheng2020, phuong2019} over vertically and horizontally partitioned datasets\footnote{For horizontally partitioned datasets, participants hold subsets of the samples with the same features, while for vertically partitioned datasets, participants hold the same samples with different features \cite{yang2019}}. 

Like the above methods, gradient tree boosting \cite{friedman2001} which is one of the most popular machine learning methods has also received considerable attention due to its effectiveness in a wide range of application areas such as fraud detection \cite{minastireanu2019}, feature selection \cite{punmiya2019} and product recommendation \cite{wang2016}. Efforts to address the privacy concerns for gradient tree boosting in multi-party setting are presented in \cite{li2020, ong2020, liu2019, cheng2019, feng2019, fang2020}. The datasets in \cite{li2020, ong2020, liu2019} are horizontally partitioned while the datasets are vertically partitioned in \cite{cheng2019, feng2019, fang2020}.

In this work, we focus on the latter dataset partitioning. The current privacy preservation efforts proposed for the multi-party gradient tree boosting method with vertically partitioned datasets have a number of limitations. In \cite{fang2020}, the proposed scheme is not scalable, it is limited to only two collaborating parties. And in \cite{cheng2019, feng2019}, intermediate information gets revealed during the model training. Thus, designing a scalable and yet secure gradient tree boosting scheme has remained open for investigation, and hence we intend to answer the question, how to construct a scalable and secure XGBoost \cite{chen2016} over vertically partitioned datasets in this work.

Apart from the high memory usage challenge, the secure XGBoost model training requires complicated computation primitives such as $division \text{ and } argmax$ \cite{fang2020}. To address these challenges and build a scalable but secure XGBoost over vertically partitioned datasets, we propose the SSXGB framework that securely outsources and performs the complicated computations in encrypted form. The key idea is to allow the participants to jointly train a model by sharing their encrypted information with a server that in turn collaborates with a second server to securely perform further computations to complete the generation of the model. Specifically, we present an additive homomorphic encryption (HE) scheme that provides the addition (\texttt{Add}) and subtraction (\texttt{Sub}) primitives. We also present sub-protocols designed to provide additional primitives such as the multiplication (\texttt{Mult}) and comparisons. Next, we propose new sub-protocols based on the HE scheme for the division (\texttt{Div}) and \texttt{argmax} primitives. We employ the secure computation primitives to build the scalable and secure XGBoost model. Then, we present a secure prediction algorithm for predictions based on the trained model. We present the analysis of our framework and its implementation using real-world datasets. A summary of our contributions are presented as follows:

\begin{itemize}
  \item We propose sub-protocols based on an additive HE scheme used to perform primitive secure operations during a machine learning task. The sub-protocols are collaboratively executed by two non-colluding servers. 
  \item We design a novel scalable and privacy-preserving multi-party XGBoost training algorithm and a corresponding prediction algorithm. The algorithms are constructed under the semi-honest security assumption and there is no limit on the number of participants involved.
  \item We conduct experiments using real-world datasets to demonstrate the effectiveness and efficiency of our proposed framework. 
\end{itemize}

The rest of the paper is organized as follows. In section \ref{sec:related_work}, we present the related works. Section \ref{sec:preliminaries} contains the preliminary concepts. In section \ref{sec:premitives}, we present our proposed HE sub-protocols for non-linear operations. We present the overview of the proposed SSXGB framework in section \ref{sec:overview}. Sections \ref{sec:training} and \ref{sec:prediction} present the secure training and prediction algorithms of the SSXGB framework. In section\ref{sec:analysis}, we present theoretical security and communication analysis. Performance evaluation is presented in section \ref{sec:performance} and section \ref{sec:conclusion} concludes the paper.  

\section{Related work}\label{sec:related_work}
Recently, efforts devoted to multi-party machine learning researches have shown a huge potential in addressing the training data scarcity problem while preserving the data privacy \cite{gong2020, aono2016, aono2017, shokri2015}. However, the majority of the works focus on linear machine learning models. Little effort has been invested in researching multi-party gradient tree boosting models. 
Currently, a multi-party gradient tree boosting framework can be categorized as a \textit{horizontal} or \textit{vertical} framework, depending on how its dataset is partitioned amongst the collaborating participants.

\subsection{Horizontal Multi-party Gradient Tree Boosting Frameworks}
In horizontal multi-party gradient tree boosting frameworks, dataset features are shared amongst the collaborating participants. Several works have adopted this approach. In \cite{ong2020}, Ong \textit{et al.} designed a multi-party gradient tree boosting framework in which the participants exchange adaptive histogram representation of their data during model learning. Liu \textit{et al.} combined secret sharing with homomorphic encryption to prevent participants from dropping out and securely aggregate their gradients during XGBoost training \cite{liu2019}. \cite{leung2020} employed an oblivious algorithm to prevent privacy violations at hardware enclaves during learning of a multi-party gradient tree boosting model. In \cite{yang2019tradeoff}, Yang \textit{et al.} designed a multi-party tree boosting framework with anomaly detection from extremely unbalanced datasets. \cite{wang2020} designed secure training and prediction frameworks for multi-party gradient tree boosting. A secret share scheme is employed for the secure training while a key agreement scheme and an identity-based encryption and signature scheme are employed for the secure prediction framework. Unlike the above frameworks, our work focuses on vertically partitioned datasets. 

\subsection{Vertical Multi-party Gradient Tree Boosting Frameworks}
In vertical multi-party gradient tree boosting frameworks, sets of samples are shared amongst the collaborating participants. Several existing efforts have focused on addressing concerns in this setting. In \cite{cheng2019}, Cheng \textit{et al.} designed a lossless privacy-preserving multi-party gradient tree boosting framework using a homomorphic encryption scheme. The framework achieves the same accuracy as the non-federated gradient tree boosting frameworks. However, it reveals the intermediate parameters during the training process which can lead to privacy violations. In \cite{fang2020, feng2019}, the authors proposed secure training and prediction frameworks for privacy-preserving multi-party gradient tree boosting. However, their schemes are unscalable, i.e., they are limited to two parties. In contrast, in our work, we proposed a privacy-preserving multi-party gradient tree boosting framework that is scalable and does not expose the intermediate parameters. 

\section{Preliminaries}\label{sec:preliminaries}
This section summarizes the gradient tree boosting framework, XGBoost, and the cryptographic foundations used to construct our proposed privacy-preserving multi-party gradient tree boosting framework.

\subsection{XGBoost}
XGBoost is an implementation of gradient tree boosting. It iteratively minimizes the loss sum of all the samples in an additive manner \cite{chen2016}. Normally, the loss function $l(y,\hat{y})$ is defined to minimize the difference between the predicted and the true values. In XGBoost, to obtain the predicted values, regression trees over a given dataset are used. The trees are greedily added to one another after every iteration \cite{chen2016, cheng2019, fang2020}.

To fit a tree, for each sample $i$, the algorithm generates the first order derivative as $g_i$ and the second order derivative as $h_i$, i.e.,
\begin{equation}
  g_i = \partial_{\hat{y_i}^{(t-1)}}l(y_i,\hat{y_i}^{(t-1)}) \text{ and } h_i = \partial^{2}_{\hat{y_i}^{(t-1)}}l(y_i, \hat{y_i}^{(t-1)})
\end{equation}
where $\hat{y_i}^{(t-1)}$ denotes the predicted value for the last iteration. 

The sum of $g_i$ and $h_i$ for a nodes' instance set $I_j$ can be computed as:
\begin{equation}
  G_j = \sum_{i\in I_j}g_i \text{ and } H_j = \sum_{i\in I_j}h_i.
\end{equation} 

The optimal weight $w^{\ast}_{j}$ for the leaf node is obtained as:
\begin{equation}\label{eqn:w}
  w^{\ast}_{j} = -\frac{\sum_{i\in I_j}g_i}{\sum_{i\in I_j}h_i + \lambda}
\end{equation}
where $\lambda$ is the regularizer for the leaf weight.

At each iteration, i.e., during the construction of each tree, Equation \ref{eqn:split}, is iteratively used for split decisions at each intermediate node from depth 0 until the maximum depth is reached. 
\begin{equation}\label{eqn:split}
  {L}_{split} = \frac{1}{2}[\frac{(\sum_{i\in I_L} g_i)^2}{\sum_{i\in I_L}h_i + \lambda} + \frac{(\sum_{i\in I_R} g_i)^2}{\sum_{i\in I_R} h_i + \lambda} - \frac{(\sum_{i\in I} g_i)^2}{\sum_{i\in I} h_i + \lambda}] - \gamma
\end{equation}
where $ \gamma$ is the regularizer for the leaf number, and $I_L \text{ and } I_R$ are the instance set for the left and right child nodes. Thus, $G_L = \sum_{i\in I_L} g_i$ and $H_L = \sum_{i\in I_L} h_i$ denote the sum of $g_i$ and $h_i$ for the left child node instance space, and $G_R = \sum_{i\in I_R} g_i$ and $H_R = \sum_{i\in I_R} h_i$ denote the sum of $g_i$ and $h_i$ for the right child node instance space.

\subsection{Homomorphic Encryption (HE)}

In this work, we adopt the BCP Scheme \cite{bresson2003, peter2013}, which is an additive homomorphic encryption scheme. The scheme comprises the following algorithms:

\textbf{(1) Setup($k$):} For a given security parameter $k$, and two large primes $p$ and $q$ of length $k$ bits, the algorithm generates the public parameters ($pp$) and the master key ($mk$) as follows. First, it computes $N=pq$. It then randomly chooses $g\in \mathbb{Z}^{*}_{N^2}$ of order $pp^\prime qq^\prime$ (s.t. $g^{p^\prime q^\prime} = 1+kN$ for $k\in [1, N-1]$), where $p^\prime=\frac{p-1}{2}$ and $q^\prime = \frac{q-1}{2}$. The algorithm outputs $pp$ as ($N, g, k$) and $mk$ as ($p^\prime, q^\prime$).

\textbf{(2) KeyGen($pp$):} Generates the public-secret key pairs for users. To generate the key pair for a user $i$, the algorithm randomly picks $a_i \in \mathbb{Z}_{N^2}$, and outputs the public key ($pk_i$) as $h_i = g^{a_i}\text{ mod }N^2$ and the secret key ($sk_i$) as $a_i$.

\textbf{(3) Enc($pp,pk_i,m$):} Encrypts the message $m \in \mathbb{Z}_N$ under the public key $pk_i$. To encrypt $m$, the algorithm randomly chooses $r \in \mathbb{Z}_{N^2}$ and outputs the ciphertext ($CT$) as ($A,B$), where
\[ A = g^r\text{mod }N^2 \text{ and } B = h_{i}^{r}(1+mN)\text{mod }N^2. \]

\textbf{(4) Dec($pp,sk_i, CT$):} Recovers the message $m$ from $CT = (A,B)$ using the corresponding $sk_i = a_i$. The recovery is performed as follows:
\begin{equation}
  m = \frac{B/(A^{a_i})-1 \text{ mod }N^2}{N}.
\end{equation}
Note that the above recovery is successful only for the $pk_i$-$sk_i$ pair. 

\textbf{(5) mDec($pp,pk_i,mk,CT$):} Recovers any properly created ciphertext using the master key, i.e., the algorithm can decrypt $CT$ encrypted under any users' public key $pk_i$ (so long as $pk_i$ is legitimate). For the decryption to proceed, first $a \text{ mod }N$ and $r \text{ mod }N$ are computed as:
\begin{equation}
  a \text{ mod }N = \frac{h^{p^\prime q^\prime}-1 \text{ mod }N^2}{N}.k^{-1}\text{ mod } N
\end{equation}
and 
\begin{equation}
  r \text{ mod }N = \frac{A^{p^\prime q^\prime}-1 \text{ mod }N^2}{N}.k^{-1}\text{ mod }N,
\end{equation}
where $k^{-1}$ is the inverse of $k \text{ mod } N$.
$m$ is recovered from $CT$ as:
\begin{equation}
  m = \frac{(B/(g^\gamma))^{p^\prime q^\prime}-1 \text{ mod }N^2}{N}.\delta \text{ mod }N,
\end{equation}
where $\delta$ is the inverse of $p^\prime q^\prime \text{ mod } N$ and $\gamma:= ar \text{ mod }N$.

\subsection{BCP Sub-protocols}
To perform arithmetic operations homomophically, \cite{peter2013, bresson2003, liu2016} proposed the following subprotocols. 

\textbf{(a)} \texttt{KeyProd}: The sub-protocol transforms the encryptions under the different user public keys $pk_1, \cdots, pk_n$ to encryptions under a joint public key $pk_{\Sigma}:=\prod_{i=1}^{n}pk_i$. The transformation is an interactive process that involves two non-colluding servers (see \cite{peter2013} for the details). The encryption under the joint public key $pk_{\Sigma}$ can only be decrypted using the sum of all the user secret keys $sk_{\Sigma} = \sum_{1}^{n} sk_i$ or the master key.

\textbf{(b)} \texttt{Add}: Returns the encrypted sum of two encrypted messages. Suppose, two messages $m_1$ and $m_2$ are encrypted as \textlbrackdbl$m_1\text{\textrbrackdbl }$ and \textlbrackdbl$m_2$\textrbrackdbl, respectively. \textlbrackdbl $.\text{\textrbrackdbl }$ denotes an encryption operation under the joint public key in our case. The \texttt{Add} sub-protocol sums the two ciphertexts as \textlbrackdbl$m_1+m_2$\textrbrackdbl. The fact that the BCP HE scheme achieves its additive nature under the same public key straightaway simplifies the \texttt{Add} sub-protocol. The encryptions under different user public keys can first be transformed to be under the same public key using the \texttt{KeyProd} sub-protocol, then followed by their addition. Thus, the sum of two encryptions ($A,B$) and ($A^\prime, B^\prime$) under the same public key can be computed as (see \cite{peter2013} for the details):
\[
  (\bar{A}, \bar{B}) \leftarrow (A.A^\prime \text{mod }N^2, B.B^\prime \text{mod }N^2).
\]

\textbf{(c)} \texttt{Mult}: The \texttt{Mult} sub-protocol returns the encrypted product of two ciphertexts. The process involves interaction between two non-colluding servers (see \cite{peter2013} for the details). Thus, given two ciphertexts \textlbrackdbl $m_1$\textrbrackdbl and \textlbrackdbl $m_2$\textrbrackdbl, the \texttt{Mult} sub-protocol returns \textlbrackdbl $m_1\times m_2$\textrbrackdbl.

\textbf{(d)} \texttt{TransDec}: The \texttt{TransDec} sub-protocol does the opposite of the \texttt{KeyProd} sub-protocol. It transforms the encryptions under the joint public key $pk_{\Sigma}$ to encryptions under the user public keys $pk_1, \cdots, pk_n$. The transformation process involves interactions between two non-colluding servers (see \cite{peter2013} for the details). 

\textbf{(e)} \texttt{Neg}: The \texttt{Neg} sub-protocol negates an encrypted message. For example, given an encryption of a message $m$ as \textlbrackdbl $m$\textrbrackdbl, the \texttt{Neg} subprotocol transforms it to \textlbrackdbl $-m\text{\textrbrackdbl }$ as (see \cite{bresson2003, liu2016} for the details):
\[
\big{(}g^{(N-1).r} \text{mod }N^2, h^{(N-1).r}(1+m.N)^{N-1} \text{mod }N^2\big{)} = \text{\textlbrackdbl} -m\text{\textrbrackdbl} .
\]

\textbf{(f)} \texttt{Exp}: Using the same principles as in (e) above, the \texttt{Exp} sub-protocol returns the product of an encrypted message \textlbrackdbl $m\text{\textrbrackdbl }$ and a constant $k$ as \textlbrackdbl $km$ \textrbrackdbl. It can be computed as below:
\[
\big{(}g^{(N+k).r} \text{mod }N^2, h^{(N+k).r}(1+m.N)^{N+k} \text{mod }N^2\big{)} = \text{\textlbrackdbl} km\text{\textrbrackdbl}.
\]
The correctness is similar to the \texttt{Neg} sub-protocol. 

\textbf{(g)} \texttt{Sub}: The \texttt{Sub} sub-protocol returns the difference between two ciphertexts. For example, given $m_1$ and $m_2$ encrypted as \textlbrackdbl $m_1$\text{\textrbrackdbl } and \textlbrackdbl $m_2$\textrbrackdbl, respectively, the \texttt{Sub} sub-protocol returns \textlbrackdbl $m_1 - m_2$\textrbrackdbl. We describe the sub-protocol as follows:
First, \textlbrackdbl $m_2$\text{\textrbrackdbl } is negated using the \texttt{Neg} sub-protocol. Then, the \texttt{Add} is used to complete the process. Thus,
\[
\text{\textlbrackdbl} m_1 - m_2\text{\textrbrackdbl}  = \text{\texttt{Add}}(\text{\textlbrackdbl} m_1\text{\textrbrackdbl},\text{\texttt{Neg}}(\text{\textlbrackdbl} m_2\text{\textrbrackdbl})).
\]

\textbf{(h)} \texttt{LGT}: The less than or greater than (LGT) sub-protocol shows the relationship between two ciphertexts, i.e., $\text{\textlbrackdbl}m_1 \text{\textrbrackdbl} \geq \text{\textlbrackdbl}m_2 \text{\textrbrackdbl }$ or $\text{\textlbrackdbl} m_1\text{\textrbrackdbl} <\text{\textlbrackdbl}m_2\text{\textrbrackdbl}$. The sub-protocol returns $1$ if $\text{\textlbrackdbl} m_1\text{\textrbrackdbl} <\text{\textlbrackdbl}m_2\text{\textrbrackdbl}$, it returns $0$ otherwise. It is an adaptation of the SLT protocol in \cite{liu2016}. A detailed description is presented in Appendix \ref{app:lgt}.

\section{Proposed Computation Primitives}\label{sec:premitives}
\begin{figure*}[!ht]
\centering
\includegraphics[width=6in]{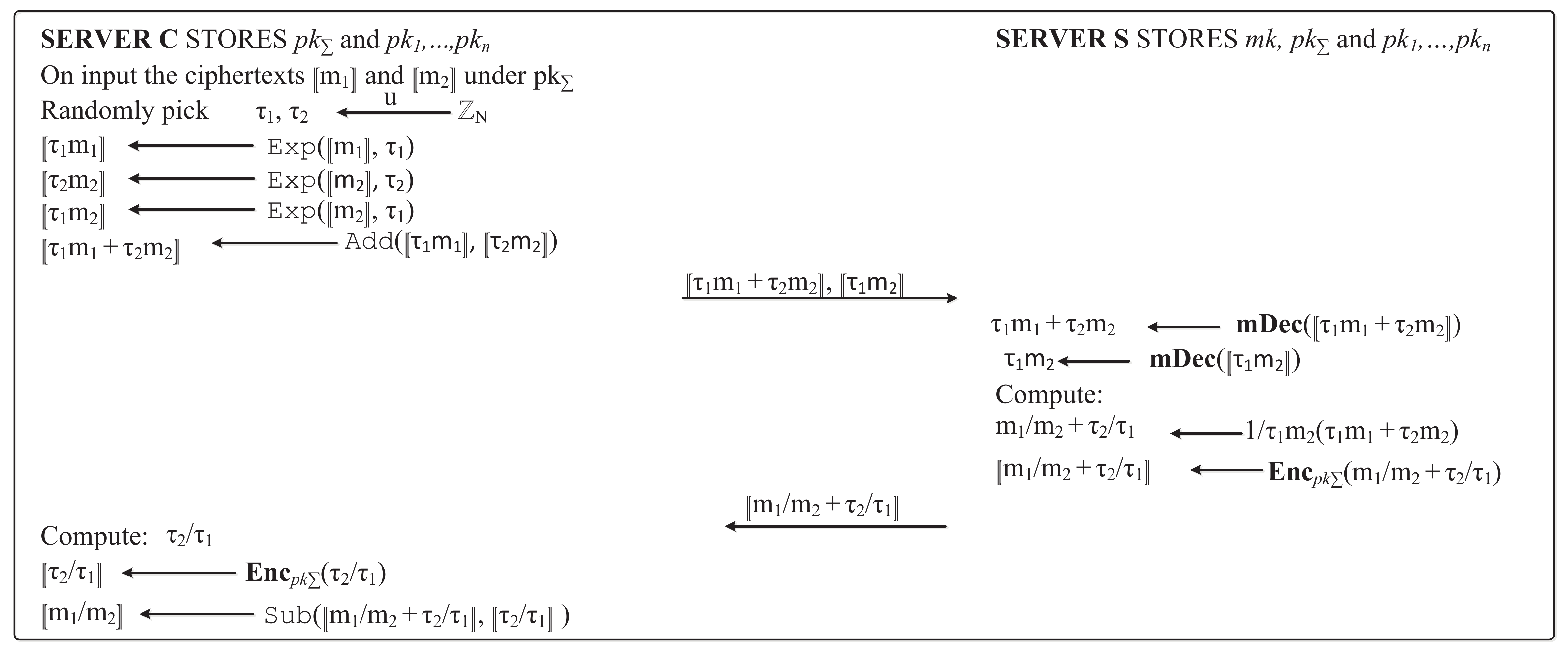}
\caption{Illustration of the \texttt{Div} sub-protocol}
\label{fig:div}
\end{figure*}

Building gradient tree boosting algorithms requires more complicated computation primitives such the \textit{division} and \textit{argmax}. The division sub-protocol based on BCP scheme proposed in \cite{liu2016} is inefficient and unsuitable for our setting. Thus, we propose two sub-protocols \texttt{Div} and \texttt{Sargmax} based on the BCP HE scheme to perform the \textit{division} and \textit{argmax} operations, respectively. 

\subsubsection{\texttt{Div}}
The \texttt{Div} sub-protocol outputs the encrypted division of two ciphertexts. Given two ciphertexts $\text{\textlbrackdbl} m_1\text{\textrbrackdbl} $ and $\text{\textlbrackdbl} m_2\text{\textrbrackdbl}$, the \texttt{Div} sub-protocol returns $\text{\textlbrackdbl} m_1 \div m_2\text{\textrbrackdbl}$, $m_1$ being the nominator and $m_2$ the denominator. The protocol is run interactively between two non-colluding servers, say server C and server S as illustrated in Figure \ref{fig:div}. 

Using the \texttt{Exp} sub-protocol, the server C first masks the ciphertexts $\text{\textlbrackdbl} m_1\text{\textrbrackdbl} $ and $\text{\textlbrackdbl} m_2\text{\textrbrackdbl}$ as
$\text{\textlbrackdbl} \tau_1 m_1\text{\textrbrackdbl}$ and $\text{\textlbrackdbl} \tau_2 m_2\text{\textrbrackdbl}$, respectively. Where $\tau_1, \tau_2 \in \mathbb{Z}_N$. The server C then sends $\text{\textlbrackdbl} \tau_1 m_1 + \tau_2 m_2\text{\textrbrackdbl} $ and $\text{\textlbrackdbl} \tau_1 m_2\text{\textrbrackdbl}$ to the server S. The two ciphertexts are decrypted by the server S. In plaintext, the server S performs $\frac{1}{\tau_1 m_2}\times (\tau_1 m_1 + \tau_2 m_2)$ and encrypts the result and send it back to the server C. The server C extracts $\text{\textlbrackdbl} \frac{m_1}{m_2}\text{\textrbrackdbl}$ by subtracting $\text{\textlbrackdbl} \frac{\tau_2}{\tau_1}\text{\textrbrackdbl}$ out of the result received from the server S. See Appendix \ref{app:div} for the proof of correctness.

\subsubsection{\texttt{Sargmax}}
The \texttt{Sargmax} sub-protocol returns the arguments of the maximum value. In our case, the maximum value returned is in encrypted form. The maximum encrypted value is obtained using the \texttt{LGT} sub-protocol. Once the maximum value is obtained, its associated arguments are returned as the \texttt{Sargmax} sub-protocol's result. The details are shown in Algorithm \ref{alg:sargmax}.

\SetNlSty{textbf}{}{:}
{\SetAlgoNoLine
  \begin{algorithm}[!t]
%
    \textbf{function} {Sargmax}(dict)\\
    
    \Indp //Input: dict -is a dictionary of encrypted values\\
    max=None\\
    \For{key \textbf{in} dict.keys()}{\eIf{max==None}{max = dict[key]}{
    //Securely compare the encrypted values\\
    \texttt{LGT}(max, dict[key])\\
    \eIf{max$>$dict[key]}{max=max}{max=dict[key]}
    }
    }
    \textbf{return} key\\
    \Indm \textbf{end}
     
    \caption{Secure \texttt{argmax} Computation Algorithm}\label{alg:sargmax}
  \end{algorithm}}

In this section, we first describe the involved entities, followed by the workflow of our proposed framework.

\subsection{Entities of the Framework}
\begin{figure}[!ht]
\centering
\includegraphics[width=3.5in]{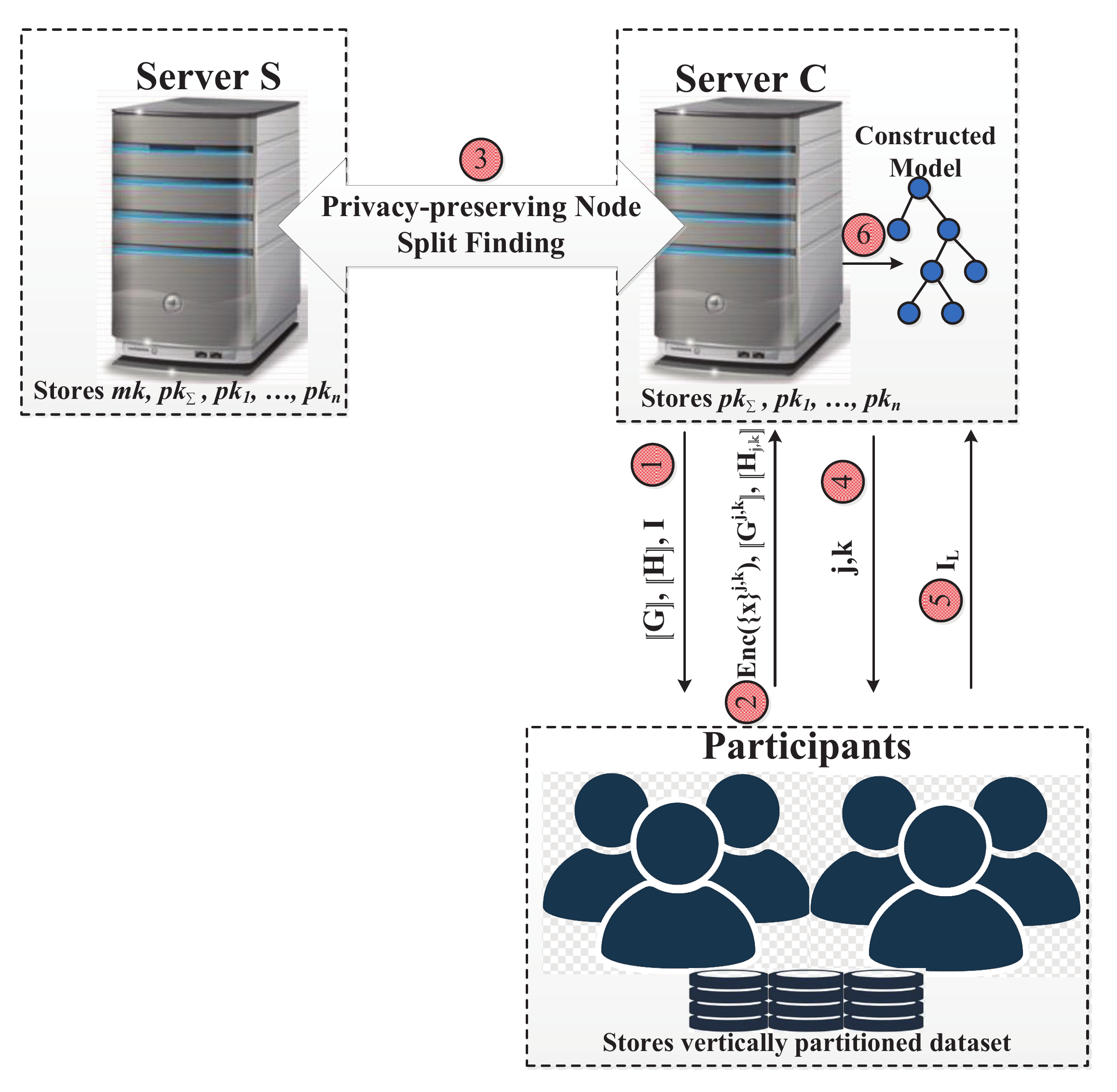}
\caption{Entities of our proposed framework}
\label{fig:arch}
\end{figure}
Our proposed framework comprises three types of entities: a set of participants, and two servers S and C as illustrated in Figure \ref{fig:arch}.

\textbf{Participants:} Participants are volunteers willing to take part in multi-party gradient tree boosting model learning. In this work, each participant holds a portion of a vertically partitioned dataset. We refer to the participant holding the label feature as Label Bearing Participant (LBP). There is only one LBP. Each participant only interacts with Server C. Additionally, each participants holds a public-private key pair. 

\textbf{Server S:} S holds $pk_{\Sigma}$, $mk$ and $pk_1, \cdots, pk_n$. Thus, S can decrypt any legitimately encrypted message. However, S only communicates directly with server C, i.e., it does not directly access the participants' data. It mainly helps with decrypting masked data from server C. We assume that S is \textit{honest-but-curious} \cite{mcmahan2017, truex2019, bonawitz2017}. 

\textbf{Server C:} C holds $pk_{\Sigma}$ and $pk_1, \cdots, pk_n$. C directly communicates with all the other entities. It has access to the instance space and encrypted intermediate parameters received from the participants. C and S then collaboratively perform computations on the received parameters to build the model. All the data from C to S are masked to prevent S from observing the actual contents of the data. We also assume that C is \textit{honest-but-curious}. An example of C is a cloud server.

\subsection{Workflow of Our Proposed Scheme}
The general workflow of our proposed SSXGB training is shown in Algorithm \ref{alg:workflow}. As shown in Algorithm \ref{alg:workflow}, three major protocols namely: L{\scriptsize BP}X{\scriptsize GB}T{\scriptsize RAIN}, SB{\scriptsize UILD}T{\scriptsize REE} and SP{\scriptsize RED}T{\scriptsize REE} are invoked during the model learning. To prevent inference attacks on label information, we adopt the second proposal of \cite{cheng2019}, where the first tree is built by the LBP. The L{\scriptsize BP}X{\scriptsize GB}T{\scriptsize RAIN} protocol is thus executed by the LBP for the above purpose.

\SetNlSty{textbf}{}{:}
{\SetAlgoNoLine
  \begin{algorithm}[!t]
%
    \textbf{function} S{\scriptsize XGB}T{\scriptsize RAIN}(X, Y)\\
    
    \Indp //Input: X:$\{\text{X}^i\}_{i=1}^{n}$ is an aggregation of X$^i$ from $n$ participants.\\
    //Input: Y is the label borne by the LBP\\
    \For{t=$\text{1},\cdots,\text{T}$}{
    \eIf{TreeList==Empty}{L{\scriptsize BP}X{\scriptsize GB}T{\scriptsize RAIN}($\text{X}^{lbp}\text{, Y}$)}{
    Compute: \textlbrackdbl G$_{t-1}$\textrbrackdbl:$\sum_{i}(\text{\textlbrackdbl}g_{t-1}\text{\textrbrackdbl})$ and
    \textlbrackdbl H$_{t-1}$\textrbrackdbl: $\sum_{i}(\text{\textlbrackdbl}h_{t-1}\text{\textrbrackdbl})$ \\
    //Construct a tree using \textlbrackdbl G$_{t-1}$\textrbrackdbl and \textlbrackdbl H$_{t-1}$\textrbrackdbl \\
    $F_t = \text{SB{\scriptsize UILD}T{\scriptsize REE}}$($\text{\textlbrackdbl}G_{t-1}\text{\textrbrackdbl}, \text{\textlbrackdbl}H_{t-1}$\text{\textrbrackdbl})\\
    //Predict using the current tree\\
    \textlbrackdbl $\hat{Y}_t\text{\textrbrackdbl} = \text{SP{\scriptsize RED}T{\scriptsize REE}}$($F_t, X$)\\
    T{\scriptsize REE}L{\scriptsize IST}$.\text{append}(F_t)$\\
    \textlbrackdbl $\hat{Y}\text{\textrbrackdbl = \textlbrackdbl} \hat{Y}\text{\textrbrackdbl + \textlbrackdbl} \hat{Y}_t\text{\textrbrackdbl}$
    }}
    \textbf{return} T{\scriptsize REE}L{\scriptsize IST}\\
    \Indm \textbf{end}
     
    \caption{Scalable and Secure XGBoost Training}\label{alg:workflow}
  \end{algorithm}}

Once the L{\scriptsize BP}X{\scriptsize GB}T{\scriptsize RAIN} protocol is executed, the returned parameters are encrypted and sent to server C for the rest of the participants to join the process. The two protocols SB{\scriptsize UILD}T{\scriptsize REE} and SP{\scriptsize RED}T{\scriptsize REE} are then iteratively executed by all the participants and the servers to complete the model learning process. At each iteration $t$, the SB{\scriptsize UILD}T{\scriptsize REE} is invoked to securely build a tree, while the SP{\scriptsize RED}T{\scriptsize REE} is invoked to make predictions using the built tree at $t$. Finally, the protocol returns a trained model T{\scriptsize REE}L{\scriptsize IST}. The details are presented in the subsequent sections.

\section{Scalable and Secure Multi-Party XGBoost Building}\label{sec:training}
This section presents the building of our proposed SSXGB model over vertically partitioned dataset. We specify that all participants bear distinct sets of data features. The LBP bears the label feature. We also assume that the participants have the same samples. We emphasize that server C operates only on encrypted data and does not have direct access to the participants' data, including the label.

\SetNlSty{textbf}{}{:}
{\SetAlgoNoLine
  \begin{algorithm}
    \textbf{function} L{\scriptsize BP}X{\scriptsize GB}T{\scriptsize RAIN}($\text{X}^{lbp}\text{, Y}$)\\
    
    \Indp //Input: X$^{lbp}$:$\{x^{i,j}_{M\times \mathcal{N}_{lbp}}\}$ where $\mathcal{N}_{lbp}$ is the number of features borne by the LBP \\
    //Input: Y:$\{y\}^{M}_{i=1}$ is the label\\
    //Compute the base score\\
    F$_0=\text{C{\scriptsize OMPUTE}B{\scriptsize ASE}S{\scriptsize CORE}(Y)}$\\
    T{\scriptsize REE}L{\scriptsize IST} = []\\
    //Initial prediction\\
    $\hat{\text{Y}}=F_0$\\
    Compute: G$_0$:$\sum_{i}g_0$ and H$_0$:$\sum_{i}h_0$\\
    //Construct a tree using $\text{G}_{0}$ and $\text{H}_{0}$\\
    $\text{F}_1 = \text{B{\scriptsize UILD}T{\scriptsize REE}}$($\text{G}_{0}, \text{H}_{0}$)\\
    //Predict using the tree $F_1$\\
    $\hat{\text{Y}}_1 = \text{P{\scriptsize RED}T{\scriptsize REE}}$($F_1, X^{lbp}$)\\
    $\hat{\text{Y}} = \hat{\text{Y}}+\hat{\text{Y}}_1$\\
    //Encrypt the base score and the tree node values, and update the model list\\
    T{\scriptsize REE}L{\scriptsize IST}$.\text{append(Enc(F}_0\text{)}_{pk_{lbp}}, \text{Enc(F}_1\text{)}_{pk_{lbp}})$\\
    //Encrypt the label information and the updated prediction matrix\\
    Enc($\text{Y)}_{pk_{lbp}}$, Enc($\hat{\text{Y}}\text{)}_{pk_{lbp}}$\\
    \textbf{return} T{\scriptsize REE}L{\scriptsize IST}, Enc($\text{Y)}_{pk_{lbp}}$, Enc($\hat{\text{Y}}\text{)}_{pk_{lbp}}$\\
    \Indm \textbf{end}
     
    \caption{The First Tree Building by the LBP}\label{alg:lbp}
  \end{algorithm}}
  
\subsection{The First Secure Tree Building by LBP}
As stated in the previous section, to prevent participants from inferring on the label information, we adopt the proposal of \cite{cheng2019} in which the LBP builds the first tree. The L{\scriptsize BP}X{\scriptsize GB}T{\scriptsize RAIN} function shown in Algorithm \ref{alg:lbp} is invoked for that purpose. The L{\scriptsize BP}X{\scriptsize GB}T{\scriptsize RAIN} function takes as input the dataset (X$^{lbp}$, Y), where X$^{lbp}$ is the feature matrix of the LBP and Y is the label information. In other words, the LBP does not require the other participants' data to build the first tree. Since there is no collaboration in building the first tree, the sub-routines: C{\scriptsize OMPUTE}B{\scriptsize ASE}S{\scriptsize CORE}, B{\scriptsize UILD}T{\scriptsize REE} and P{\scriptsize RED}T{\scriptsize REE} for computing the base score, building the first tree and making the initial predictions, respectively, are consistent with the mechanisms of XGBoost \cite{chen2016}.

Once the first tree is constructed, LBP encrypts the base score and the node values of the tree with its public key ($pk_{lbp}$). Next, it updates the model with the encrypted base score and the tree. It also encrypts the label information and the prediction matrix with its public key. Finally, it returns T{\scriptsize REE}L{\scriptsize IST}, Enc($\text{Y)}_{pk_{lbp}}$ and Enc($\hat{\text{Y}}\text{)}_{pk_{lbp}}$. The returned parameters are sent to server C to continue with the model training.

\SetNlSty{textbf}{}{:}
{\SetAlgoNoLine
  \begin{algorithm}
    \textbf{function} \text{SB{\scriptsize UILD}SP{\scriptsize RED}T{\scriptsize REE}}($\text{\textlbrackdbl G}_{t-1}\text{\textrbrackdbl}, \text{\textlbrackdbl H}_{t-1}\text{\textrbrackdbl}$)\\
    
    \Indp //Input:- \textlbrackdbl G$_{t-1}$\textrbrackdbl:$\{\text{\textlbrackdbl}g_{t-1}\text{\textrbrackdbl }\}_{i=1}^{M}$, \textlbrackdbl H$_{t-1}$\textrbrackdbl:$\{\text{\textlbrackdbl}h_{t-1}\text{\textrbrackdbl }\}_{i=1}^{M}$\\
    \textit{/* Computed by S{\scriptsize ERVER} C */}\\
    \If{R{\scriptsize OOT}N{\scriptsize ODE}==None}{//Register the current node as the root node\\
    R{\scriptsize OOT}N{\scriptsize ODE}=C{\scriptsize URRENT}N{\scriptsize ODE}}

    \textit{/* Computed at each P{\scriptsize ARTICIPANT} $i$ */}\\
    \ForEach{feature $j$}{
    Propose split candidates $\{x\}^{j,0}, \cdots, \{x\}^{j,K}$
    }
    \ForEach{$\{x\}^{j,k}$}{
    Compute $\text{\textlbrackdbl G}^{j,k}_{t-1}\text{\textrbrackdbl}$ and $\text{\textlbrackdbl H}^{j,k}_{t-1}\text{\textrbrackdbl}$\\
      Create a lookup table and record $j,k$ and $\{x\}^{j,k}$ in the table\\
      Create a tuple ($j,k$)
    }
    Send the tuple (($j,k$), $\text{\textlbrackdbl G}^{j,k}_{t-1}\text{\textrbrackdbl}, \text{\textlbrackdbl H}^{j,k}_{t-1}$\textrbrackdbl) to S{\scriptsize ERVER} C\\
    \textit{/* Computed by S{\scriptsize ERVER} C */}\\
    $j_{opt}, k_{opt}$ = SS{\scriptsize PLIT}N{\scriptsize ODE}($\text{\textlbrackdbl G}^{j,k}_{t-1}\text{\textrbrackdbl}, \text{\textlbrackdbl H}^{j,k}_{t-1}\text{\textrbrackdbl}$, $\mathcal{T}$)\\
    
    \textit{/* Computed at the optimal P{\scriptsize ARTICIPANT} */}\\
    Receive the optimal $j_{opt}, k_{opt}$ from S{\scriptsize ERVER} C\\
    Check the lookup for $\{x\}^{j,k}$ associated with $j_{opt}, k_{opt}$\\
    Partition I based on $\{x\}^{j,k}$\\
    Record the instance space I$_L$ with S{\scriptsize ERVER} C\\
    \Indm \textbf{end}
     
    \caption{Secure B{\scriptsize UILD} and P{\scriptsize RED} T{\scriptsize REE}}\label{alg:buildpred}
  \end{algorithm}}

\subsection{Secure B{\scriptsize UILD}T{\scriptsize REE} and P{\scriptsize RED}T{\scriptsize REE}}
Once the first tree is built by the LBP and the results are returned to server C, the rest of the participants can join to continue with the model training. First, the servers C and S collaboratively transform the parameters from the LBP encrypted under the public key $pk_{lbp}$ to be under the joint public key $pk_{\Sigma}$ using the \texttt{KeyProd} sub-protocol discussed in Section \ref{sec:preliminaries}. For example, the encrypted label information Enc($\text{Y)}_{pk_{lbp}}$ is transformed to \textlbrackdbl Y\textrbrackdbl, the encrypted prediction matrix Enc($\hat{\text{Y}}\text{)}_{pk_{lbp}}$ is transformed to \textlbrackdbl$\hat{\text{Y}}$\textrbrackdbl, etc. Then server C computes the first and second order derivatives using the encrypted parameters for all the instances. Next, the joint model building resumes using the SB{\scriptsize UILD}SP{\scriptsize RED}T{\scriptsize REE} protocol shown in Algorithm \ref{alg:buildpred}.

The SB{\scriptsize UILD}SP{\scriptsize RED}T{\scriptsize REE} protocol is executed by multiple entities. The algorithm takes in as input the encrypted first and second order derivatives computed by server C. The execution begins in server C, which assigns the current node as the root node if there exists no root node. Next, each participant, including the LBP proposes split candidate values for each of their features as in \cite{chen2016}. For ease of understanding, we shall refer to all the participants including the LBP as participant $i$. 
For each split candidate, each participant $i$ computes $\text{\textlbrackdbl G}_{t-1}\text{\textrbrackdbl}, \text{\textlbrackdbl H}_{t-1}\text{\textrbrackdbl}$ which are associated with the left branch according to \cite{chen2016}. 
Each participant $i$ also creates a lookup table to record the split candidate information.
Then, for each split candidate, each participant $i$ sends the parameter tuple 
$\mathcal{T}$=(($j,k\text{)}, \text{\textlbrackdbl G}^{j,k}_{t-1}\text{\textrbrackdbl}, \text{\textlbrackdbl H}^{j,k}_{t-1}$\textrbrackdbl) 
to the server C.  

Server C then securely identifies the optimal split feature and value using the SS{\scriptsize PLIT}N{\scriptsize ODE} algorithm, and sends the result to the participant bearing the feature (optimal participant). The optimal participant then decrypts the optimal value for the optimal feature and splits the current node's instance space accordingly. Finally, the optimal participant registers the left branch instance space I$_L$ with server C after the partition.

  \SetNlSty{textbf}{}{:}
{\SetAlgoNoLine
  \begin{algorithm}[!t]
    \textbf{function} SS{\scriptsize PLIT}N{\scriptsize ODE}($\text{\textlbrackdbl}G_{t-1}\text{\textrbrackdbl}, \text{\textlbrackdbl}H_{t-1}\text{\textrbrackdbl}$, $\mathcal{T}$)\\
    
    \Indp //Input:- \textlbrackdbl G$_{t-1}$\textrbrackdbl :$\{\text{\textlbrackdbl}g_{t-1}\text{\textrbrackdbl}\}^{M}_{i=1}$, \textlbrackdbl H$_{t-1}$\textrbrackdbl:$\{\text{\textlbrackdbl}h_{t-1}\text{\textrbrackdbl}\}^{M}_{i=1}$, and $\mathcal{T}$.\\
    /*Collaboratively Computed by S{\scriptsize ERVER} S and S{\scriptsize ERVER} C*/\\
    \If{C{\scriptsize URRENT}N{\scriptsize ODE}==L{\scriptsize EAF}N{\scriptsize ODE}}{
    //Compute the weight of the leaf node\\
    $\text{\textlbrackdbl }w^*\text{\textrbrackdbl}$ = $\text{\texttt{Div}(}\Sigma_i \{\text{\textlbrackdbl}g\text{\textrbrackdbl}\}^i, (\Sigma_i \{\text{\textlbrackdbl}h\text{\textrbrackdbl}\}^i + \text{\textlbrackdbl}\lambda\text{\textrbrackdbl}))$\\
    $\{\text{\textlbrackdbl}\hat{\text{y}}\text{\textrbrackdbl}\}^i$ = $\text{\textlbrackdbl}w^*\text{\textrbrackdbl}$\\
    \textbf{return}  $\text{\textlbrackdbl}\hat{\text{Y}}\text{\textrbrackdbl}$
    }
    //Compute the C{\scriptsize URRENT}N{\scriptsize ODE}'s gain ($cgain$)\\
    $\text{\textlbrackdbl}G\text{\textrbrackdbl} = \Sigma_i \{\text{\textlbrackdbl}g\text{\textrbrackdbl}\}^i$\\
    $\text{\textlbrackdbl}H\text{\textrbrackdbl}=\Sigma_i \{\text{\textlbrackdbl}h\text{\textrbrackdbl}\}^i$\\
    $\text{\textlbrackdbl}cgain\text{\textrbrackdbl}=\text{\texttt{Div}(}\text{\texttt{Mult}(\textlbrackdbl}G\text{\textrbrackdbl},\text{\textlbrackdbl}G\text{\textrbrackdbl}\text{)},\text{(\textlbrackdbl}H\text{\textrbrackdbl}+\text{\textlbrackdbl}\lambda\text{\textrbrackdbl))}$\\
    //Initialize the gain dictionary\\
    $gainDict = \{\}$\\
    //Enumerate all the P{\scriptsize ARTICIPANTS}\\
    \For{p=$0,\cdots, P$}{
    //Enumerate all the features of a P{\scriptsize ARTICIPANT}\\
    \For{j=$0, \cdots, J$}{
    //Enumerate all the proposed thresholds\\
    \For{k=$0, \cdots, K$}{
    Receive $\text{\textlbrackdbl}G_L\text{\textrbrackdbl }$ and $\text{\textlbrackdbl}H_L\text{\textrbrackdbl }$ from a P{\scriptsize ARTICIPANT}\\
    //Compute the first derivative for the right branch\\
    $\text{\textlbrackdbl}G_R\text{\textrbrackdbl} = \text{\texttt{Sub}(\textlbrackdbl}G\text{\textrbrackdbl},\text{\textlbrackdbl}G_L\text{\textrbrackdbl}\text{)}$\\
    //Compute the second derivatives for the right branch\\
    $\text{\textlbrackdbl}H_R\text{\textrbrackdbl} = \text{\texttt{Sub}(}\text{\textlbrackdbl}H\text{\textrbrackdbl},\text{\textlbrackdbl}H_L\text{\textrbrackdbl}\text{)}$\\
    //Compute gains\\
    $\text{\textlbrackdbl}lgain\text{\textrbrackdbl} = \text{\texttt{Div}(}\text{(\textlbrackdbl}G_L\text{\textrbrackdbl}\text{)}^2,\text{(}\text{\textlbrackdbl}H_L\text{\textrbrackdbl}+\text{\textlbrackdbl}\lambda\text{\textrbrackdbl}\text{))}$\\
    $\text{\textlbrackdbl}rgain\text{\textrbrackdbl} = \text{\texttt{Div}(\textlbrackdbl }G_R\text{\textrbrackdbl )}^2,\text{(\textlbrackdbl}H_R\text{\textrbrackdbl}+\text{\textlbrackdbl}\lambda\text{\textrbrackdbl}\text{))}$\\
  $\text{\textlbrackdbl}gain\text{\textrbrackdbl} = lgain + (\text{\texttt{Sub}(}rgain,cgain\text{)})$\\
  //Update the gain dictionary\\
  $gainDict[(p,j,k)]=\text{\textlbrackdbl}gain\text{\textrbrackdbl}$
    
    }
    }
    }
    $\text{j}_{optimal}, \text{k}_{optimal}$ = $\underset{j,k}{\mathrm{Sargmax}}$($gainDict$)\\
    \textbf{return }$\text{j}_{opt}, \text{k}_{opt}$ to optimal participant p\\
    Server C receives I$_L$ from optimal participant p\\
    Server C partitions its instance space into I$_L$ and I$_R$\\
    Server C associates the C{\scriptsize URRENT}N{\scriptsize ODE} with the optimal participant p as [$\text{p}: Node^{j,k}$]\\
    \Indm \textbf{end}
     
    \caption{Securely Finding Node Split}\label{alg:split}
  \end{algorithm}}

\subsection{Secure Node Split Decision}
From Equation \ref{eqn:split}, it can be observed that the optimal split can be obtained if the values of $G_L \text{ and } H_L$, and $G_R \text{ and } H_R$ can be obtained. Hence, the secure split finding algorithm SS{\scriptsize PLIT}N{\scriptsize ODE} shown in Algorithm \ref{alg:split} takes as input the first and second encrypted derivatives $\text{\textlbrackdbl}G_{t-1}\text{\textrbrackdbl}$ and $\text{\textlbrackdbl}H_{t-1}\text{\textrbrackdbl}$ and the parameter tuple $\mathcal{T}$. In this context, we simply use $\text{\textlbrackdbl}G\text{\textrbrackdbl}$ and $\text{\textlbrackdbl}H\text{\textrbrackdbl}$ for $\text{\textlbrackdbl}G_{t-1}\text{\textrbrackdbl}$ and $\text{\textlbrackdbl}H_{t-1}\text{\textrbrackdbl}$, respectively.

First, the algorithm returns and stores the encrypted prediction matrix if the current node is a leaf node (shown in lines 4-9 of Algorithm \ref{alg:split}). Otherwise, the algorithm proceeds to securely identify the optimal score. It enumerates all the participants, their features and the proposed encrypted split candidates for each of the features. For each proposed split candidate, the algorithm computes an encrypted gain (shown in lines 21-30 of Algorithm \ref{alg:split}). The encrypted gains for all the proposed split candidates are stored in a dictionary. The algorithm then executes the \textit{Sargmax} primitive algorithm to identify the optimal feature $j_{opt}$ and the threshold value $k_{opt}$. Next, server C sends the optimal parameters $j_{opt}$ and $k_{opt}$ to the optimal participant bearing the pair $j_{opt}-k_{opt}$. Then server C receives I$_L$ from the optimal participant and uses it to split its instance space into I$_L$ and I$_R$. Server C then stores the current node and associates it with the optimal participant. 
 

\section{Secure Prediction}\label{sec:prediction}
\SetNlSty{textbf}{}{:}
{\SetAlgoNoLine
  \begin{algorithm*}
    \textbf{function} SP{\scriptsize REDICT}(T{\scriptsize REE}L{\scriptsize IST}, $x^{i,j}_{1\times \mathcal{N}}$)\\
    \Indp //Input:-T{\scriptsize REE}L{\scriptsize IST} and $x^{i,j}_{1\times \mathcal{N}}$, where $\mathcal{N}$ is the number of features for the record\\
    /* Computed by all the entities */\\
    //Client c encrypt the values of the record\\
    $\text{Enc}(x^{i,j}_{1\times \mathcal{N}})_{pk_c}$\\
    Send the encrypted record to the S{\scriptsize ERVER} C\\
    \While{True}{
    //Start from the R{\scriptsize OOT}N{\scriptsize ODE}\\
    \eIf{C{\scriptsize URRENT}N{\scriptsize ODE} $!=$ L{\scriptsize EAF}N{\scriptsize ODE}}{
    S{\scriptsize ERVER} C identifies feature $j$ for the split at C{\scriptsize URRENT}N{\scriptsize ODE}\\
    S{\scriptsize ERVER} C identifies the P{\scriptsize ARTICIPANT} $p$ bearing the feature $j$\\
    //Transform the encryption to be under the public key of the P{\scriptsize ARTICIPANT} $p$\\
    $\text{Enc(}x^{i,j}+r\text{)}_{pk_c} \leftarrow \text{Enc(}x^{i,j}\text{)}_{pk_c} + \text{Enc(r)}_{pk_c}$\\
    S{\scriptsize ERVER} C send  $\text{Enc(}x^{i,j}+r\text{)}_{pk_c}$ to S{\scriptsize ERVER} S\\
    S{\scriptsize ERVER} S decrypts $\text{Enc(}x^{i,j}+r\text{)}_{pk_c}$ using mDec\\
    S{\scriptsize ERVER} S re-encrypts as $\text{Enc(}x^{i,j}+r\text{)}_{pk_p}$ and sends to S{\scriptsize ERVER} C\\
    S{\scriptsize ERVER} C extracts $\text{Enc(}x^{i,j}\text{)}_{pk_p}$ as: $\text{Enc(}x^{i,j}\text{)}_{pk_p}\leftarrow \text{Enc(}x^{i,j}+r\text{)}_{pk_p} - \text{Enc(r)}_{pk_p}$\\
    //Collaboration with the participant bearing the optimal feature for the current node\\
    S{\scriptsize ERVER} C sends $\text{Enc(}x^{i,j}\text{)}_{pk_p}$ to the P{\scriptsize ARTICIPANT} $p$ bearing the feature $j$\\
    P{\scriptsize ARTICIPANT} $p$ decrypts $\text{Enc(}x^{i,j}\text{)}_{pk_p}$ and compares it with the threshold value at the node\\
    Based on whether $x^{i,j}$ is greater or less than the threshold, P{\scriptsize ARTICIPANT} $p$ decides on the tree branch to follow\\
    P{\scriptsize ARTICIPANT} $p$ forwards the decision to S{\scriptsize ERVER} C to continue with the process at N{\scriptsize EXT}N{\scriptsize ODE}\\
    //Update the C{\scriptsize URRENT}N{\scriptsize ODE}\\
    C{\scriptsize URRENT}N{\scriptsize ODE} = N{\scriptsize EXT}N{\scriptsize ODE}
    }
    {Return the encrypted weight \textlbrackdbl w\textrbrackdbl of the L{\scriptsize EAF}N{\scriptsize ODE}}}
     
    \caption{Secure Prediction Algorithm}\label{alg:secpred}
  \end{algorithm*}}
  
Our proposed secure prediction algorithm is collaboratively executed by all the entities as shown in Algorithm \ref{alg:secpred}. The SP{\scriptsize REDICT} algorithm takes as input the trained model T{\scriptsize REE}L{\scriptsize IST} and a record to be predicted $x^{i,j}_{1\times \mathcal{N}}$ with $N$ number of features. Suppose $x^{i,j}_{1\times \mathcal{N}}$ is held by a \textit{client c} with a public key $pk_c$. To make predictions on the record, the client first encrypts the record with his public key as $\text{Enc}(x^{i,j}_{1\times \mathcal{N}})_{pk_c}$ and sends it to server C. This prevents the privacy of the record from server C. Next, server C compares and passes the record down the tree, starting from the root node. At each node, server C identifies the participant $p$ holding the node, and the $j, k$ pair associated with the node. Server C then transforms the record value for the feature $j$ of the participant $p$'s to be under the public key of the participant $p$ as $\text{Enc}(x^{i,j})_{pk_p}$ (shown in lines 12-17 of Algorithm \ref{alg:secpred}). Server C then sends $\text{Enc}(x^{i,j})_{pk_p}$ to the participant $p$. Next, $p$ decrypts the $\text{Enc}(x^{i,j})_{pk_p}$ using his secret key and compares the value with the node's threshold in plaintext. Depending on the comparison result, the participant decides on whether to follow the left or the right child nodes of the current node and sends the decision to server C. Server C repeats the process until a leaf node is reached. Once a leaf node is reached, the algorithm returns the encrypted weight \textlbrackdbl w\textrbrackdbl of the leaf node stored in server C as its result. The final prediction result is obtained by cumulating the predictions of all the trees in T{\scriptsize REE}L{\scriptsize IST}.

\section{Analysis}\label{sec:analysis}
\subsection{Security Analysis}
We consider the semi-honest (non-colluding) model in our security analysis, i.e., we consider the scenario where all the entities adhere to the protocols but try to gather information about the other entities' input and intermediate parameters as much as they can. 

\subsubsection{Security of BCP sub-protocols}
First, we present the security analysis of all the sub-protocols used in this work. The security of the \texttt{KeyProd, Add, Mult, TransDec, Sub, Exp} and \texttt{Neg} sub-protocols under the semi-honest model have already been proven in \cite{peter2013, bresson2003}. 

\texttt{Div:} Similar to the other sub-protocols, the security of the \texttt{Div} sub-protocol is based on blinding or masking the plaintext. Given the ciphertexts (numerator and denominator), we employ the properties of the homomorphic encryption to blind the ciphertexts with random elements. These random elements serve as keys. When server S decrypts the ciphertexts, without knowing the random blinding elements, it cannot obtain any information about the nominator and the denominator. They look random. On the other hand, since server C does not have access to the decryption key, it also does not obtain any information about the ciphertexts. Note that we assume the two servers do not collude. Thus, the \texttt{Div} sub-protocol is secure in the semi-honest security model. 

\texttt{LGT} and \texttt{Sargmax:} The security of the \texttt{LGT} sub-protocol is based on the fact that server S only computes on the difference between two data values. Thus, server S obtains no information about the actual data, hence making the sub-protocol secure in the semi-honest model. Therefore, our proposed \texttt{Sargmax} sub-protocol which relies on the \texttt{LGT} is also secure in the semi-honest model.

\subsubsection{Security of SSXGB}
The security analysis of the SSXGB can be split into three parts: server S part, server C part and participant part.  

\textit{Server S part:} Server S does not have access to the sample and feature space. It only collaborates with Server C in performing computations. However, the computations performed by server S are on masked values. Thus, no information gets leaked to server S.  

\textit{Server C part:} Server C has access to the sample and feature space, and it stores leaf nodes. It also know which participant hold which feature. However, the intermediate values it has access to are encrypted, and it only performs computations on the encrypted values. Although it stores the leaf nodes, the leaf values are kept in encrypted form. Thus, no information gets leaked to server C.

\textit{Participant part:} Each participant has access to the sample space for each split. Each participant knows the intermediate nodes it holds. However, each participant does not know the actual values of the intermediate parameters apart from the LBP immediately after the construction of the first tree. The participants also do not have access to the leaf nodes. Thus, no information leaks to the participants.

Since there is no information leakage in the involved entities, the proposed SSXGB is secure in the semi-honest model.  

\subsection{Communication Overhead Analysis}
We analyze the communication overhead in terms of analyzing the communication costs associated with a single split. Here, we look at \textit{server C - participant} and \textit{server C - server S} communication costs.

\textit{server C - participant communication cost:} The \textit{server C - participant} communication cost is similar to that of \cite{cheng2019}. Given $\zeta $ as the ciphertext size, $n$ as the number of samples for the current node, $q$ as the number of samples in a bucket and $d$ as the number of features held by a participant, the communication cost can be computed as $2\times \zeta \times d \times (n/q)$. 

\textit{server C - server S communication cost:} During each split, the \textit{server C - server S} communication cost can be computed as $12 \times \zeta + (3 \times \zeta \times (n/q) \times D)$, where $D$ is the total number of features. Our proposed SSXGB experiences fairly heavy communication overhead during the collaborative computation of gains by the two servers.

\section{Performance Evaluation}\label{sec:performance}
This section presents experiments to demonstrate the effectiveness and efficiency of the proposed SSXGB. 

\subsection{Experimental Setup}
\subsubsection{Hardware and Software}
All the experiments were performed using a desktop computer having Intel Core i5-6500 CPU with 3.20GHz$\times$ 4 speed and 16GB RAM running Ubuntu 20.04 operating system. We used Python 3.8.5 and gmpy2 library for the implementation. We also used Cython 0.29.23 to speed up sections of the code. The participant and server functionalities were all executed in the same computer. Thus, latency is ignored in the experiments.

\subsubsection{Datasets}
We experimented using two datasets: Iris \cite{fisher1936} and MNIST \cite{mnist}. The Iris dataset comprises three classes of iris plants, each with 50 instances. Each instance bears the features of sepal length, sepal width, petal length and petal width. Thus, the dataset contains 150 instances with 4 features (150$\times$4).

The MNIST dataset consists of 70,000 samples of gray-scale handwritten images of digits (0-9). The training set contains 60,000 samples while the test set contains 10,000 samples. Each gray-scale sample has 28$\times$28 (784) pixels. Thus, the training set is (60,000$\times$784) and the test set is (10,000$\times$784).

\subsection{Evaluation of SSXGB}
\begin{figure*}[!t]
\begin{center}
\subfloat[Accuracy with Iris]{\includegraphics[width = 2.5in]{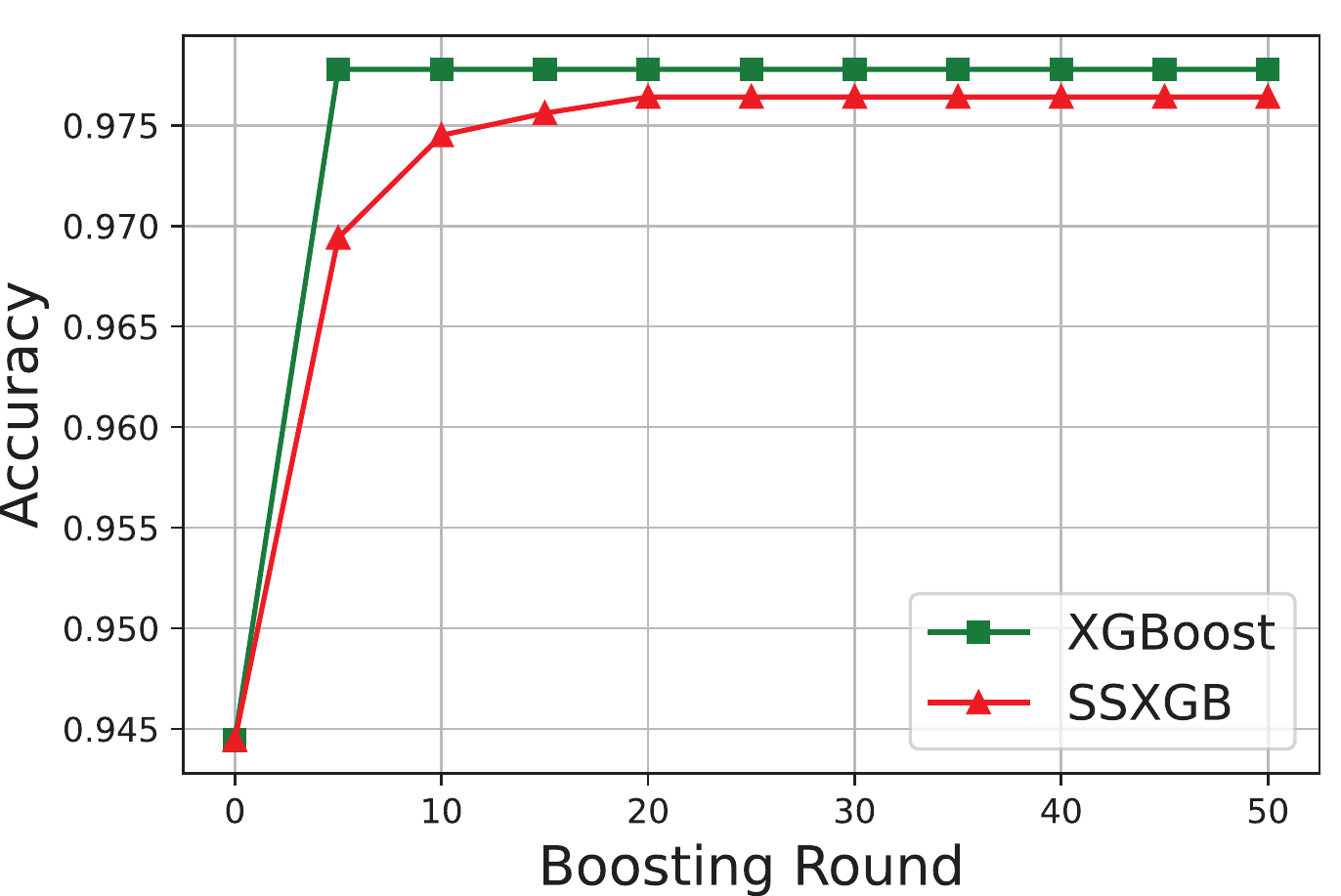}}
\subfloat[Accuracy with MNIST]{\includegraphics[width = 2.5in]{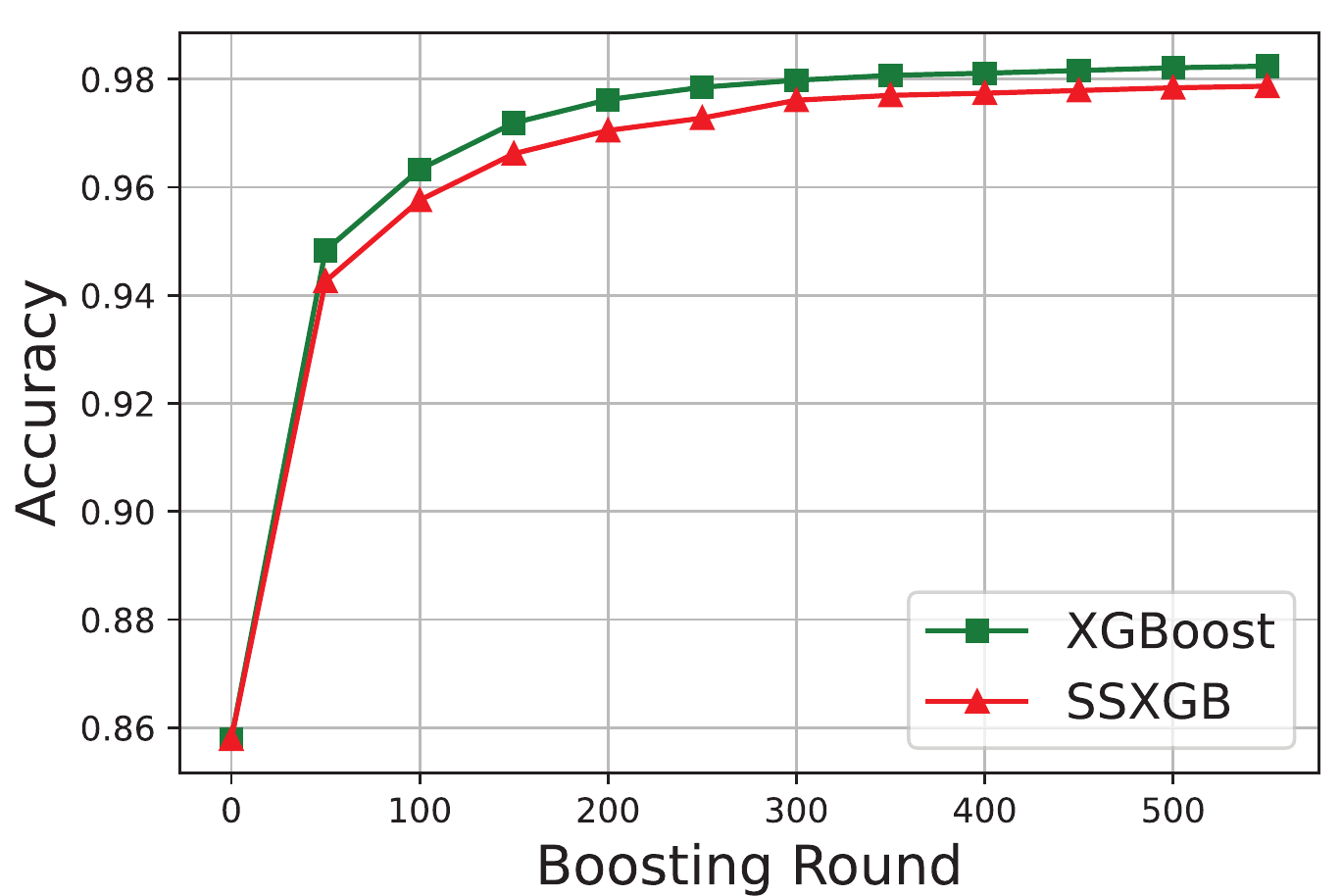}}\\
\subfloat[Loss with Iris]{\includegraphics[width = 2.5in]{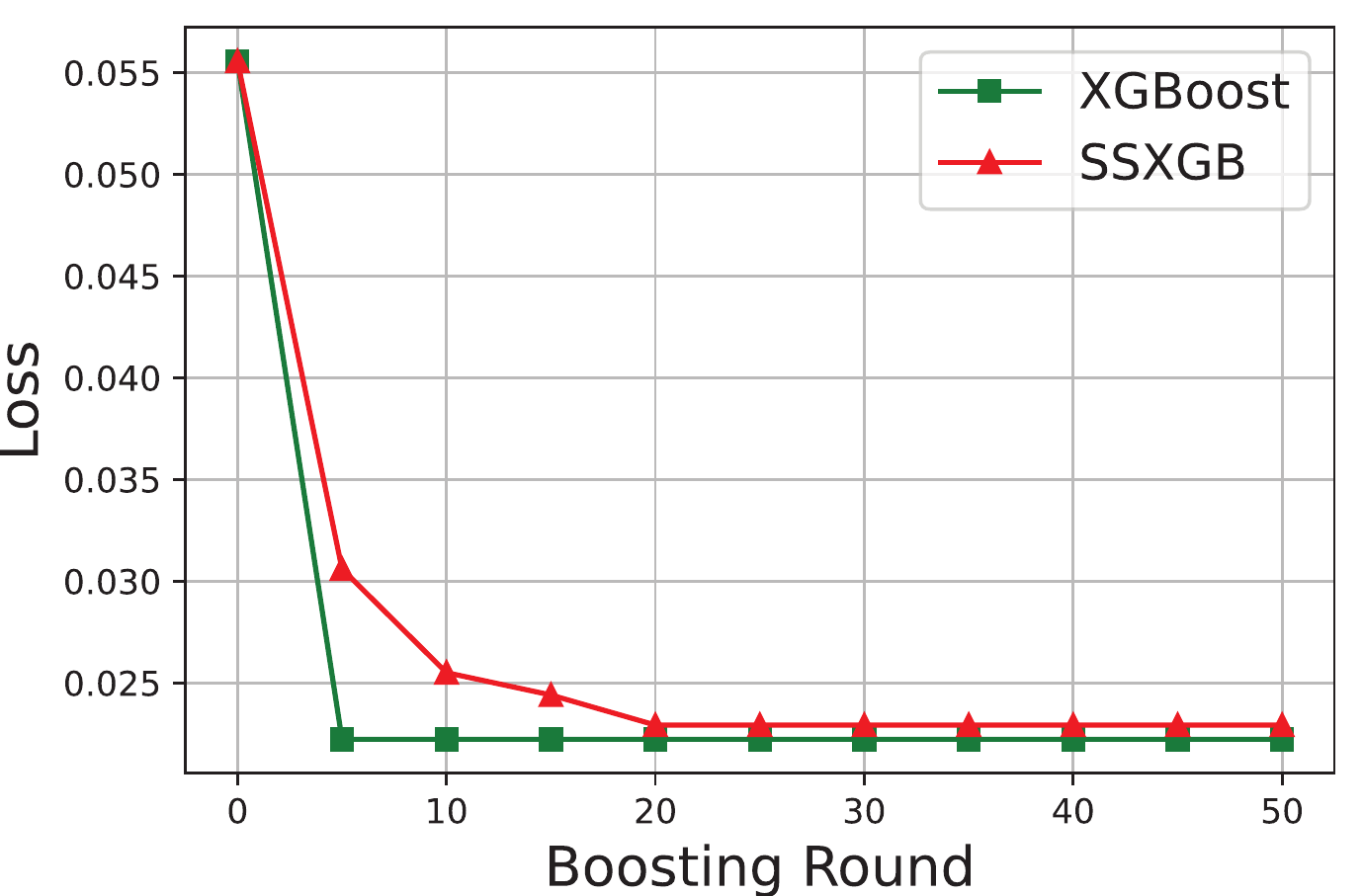}}
\subfloat[Loss with MNIST]{\includegraphics[width = 2.5in]{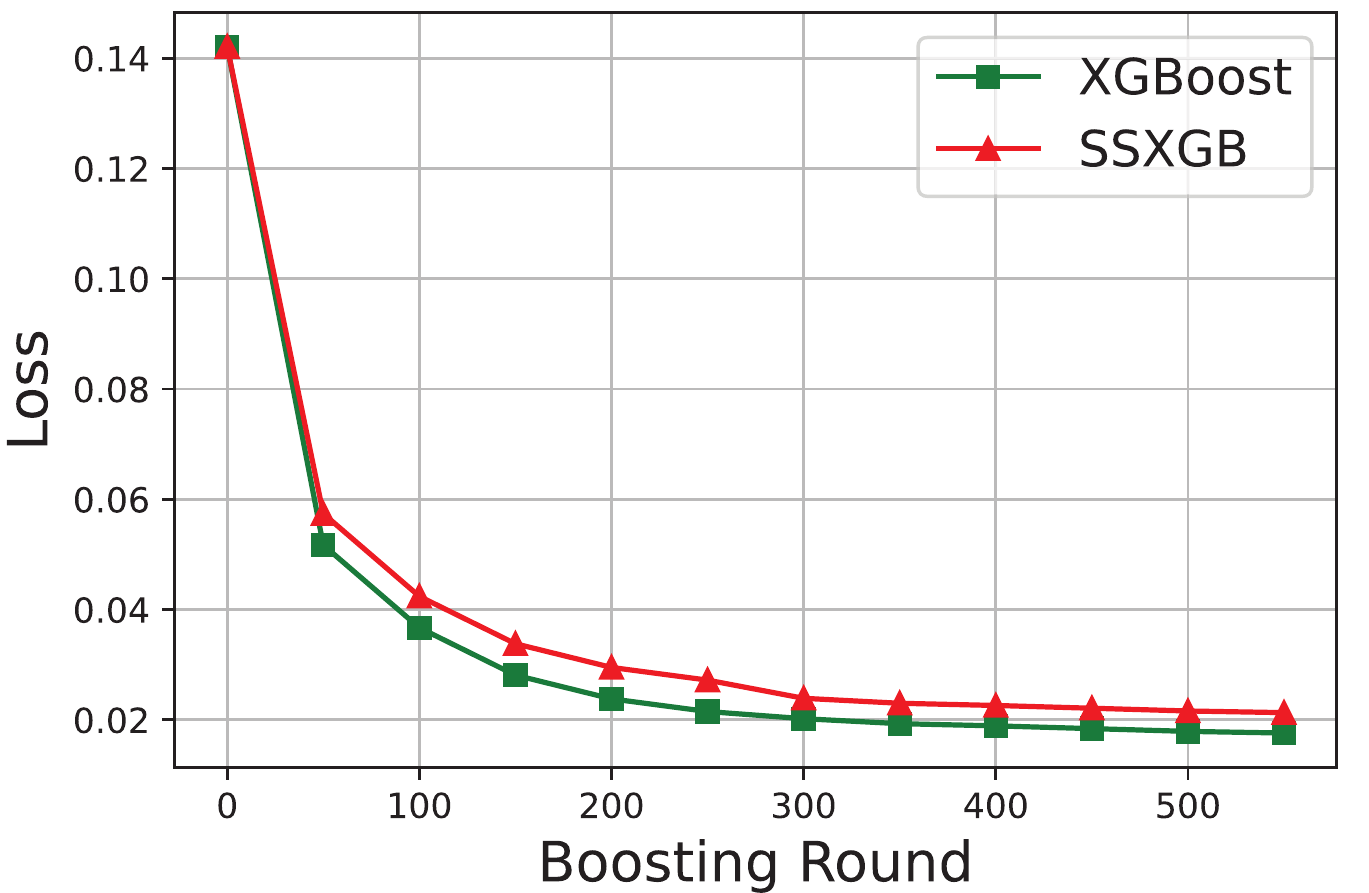}}
\caption{Training convergence with the Iris and MNIST datasets}
\label{fig:acc}
\end{center}
\end{figure*}
First, we evaluate the regression effectiveness of the proposed SSXGB with the two datasets. The effectiveness is measured in terms of regression accuracy and loss.

\textbf{Setup Configuration:} We simulated the two servers, and four (4) participants for the Iris dataset and sixteen (16) participants for the MNIST dataset. In each case, one of the participants serves as an LBP. We partitioned the datasets vertically and shared the features between the participants, i.e., for the Iris dataset, each participant held one (1) feature and for the MNIST dataset, each participant held forty-nine (49) features. Since we performed regression analysis, we considered only two classes. Thus, we take "Iris-setosa" as positive for the Iris dataset and the rest of the classes as negative. To make the dataset balanced, we synthetically generated 50 more instances of the "Iris-setosa" class. Meanwhile, for the MNIST dataset, we take the digits 0-4 as positive and 5-9 as negative. During each simulation, we set the learning rate as 0.08, and the sampling rate as 0.8 for both samples and features. We set maximum depths as 6 and 4 for the MNIST and Iris datasets experiments, respectively. We employed the approximated sigmoid function in \cite{chen2009}.

\textbf{Results:} Figure \ref{fig:acc} presents the accuracy and loss against the training round with the Iris and MNIST datasets. We can observe that the proposed SSXGB converges slowly in comparison with the XGBoost scheme, and it is slightly less effective as compared to the XGBoost, i.e., the accuracy of 0.9789 vs 0.9826 for the MNIST dataset and 0.97712 vs 0.97778 for the Iris dataset. The accuracy loss in SSXGB can be attributed to the use of the approximated sigmoid function and the restrictive nature of training the first tree. However, the slight loss in accuracy is totally acceptable.

\subsection{Privacy-Preservation Computation Overhead} 
\begin{table}[!t]
  \caption{Computation Times of BCP Algorithms and Sub-protocols}
  \label{tab:computation}

  \begin{center}
    \begin{tabular}{l|lll}
    \hline
    
    \hline
       Operation & \multicolumn{3}{|c}{Key Size (bits)}\\
       &512 & 1024 & 2048 \\
       \hline
       
       \hline
       \textbf{Enc} & 10.61 ms & 12.02 ms & 16.98 ms\\
       \textbf{Dec} & 18.27 ms & 29.13 ms & 38.64 ms\\
       \textbf{mDec} & 29.34 ms & 40.96 ms & 63.72 ms \\
       \texttt{KeyProd} & 0.02 ms & 0.05 ms & 0.11 ms\\
       \texttt{Add} & 8.86 ms & 10.23 ms & 12.61 ms\\
       \texttt{Mult} & 180.74 ms & 259.97 ms & 310.03 ms \\
       \texttt{TransDec} & 161.20 ms & 196.69 ms & 283.74 ms\\
       \texttt{Exp} & 2.04 ms & 2.32 ms & 4.01 ms\\
       \texttt{Sub} & 8.93 ms & 12.06 ms & 13.02 ms\\
       \texttt{LGT} & 120.17 ms & 143.72 ms & 171.05 ms\\
       \texttt{Div} & 140.29 ms & 183.18 ms & 251.67 ms \\
       \hline
       
       \hline
    \end{tabular}
  \end{center}
\end{table}
Since HE is crucial for privacy preservation in our proposed SSXGB, identifying a suitable configuration for the BCP scheme was paramount. Thus, we implemented the BCP scheme and measured the computation times of its algorithms and sub-protocols under different key sizes. For each operation, the experiment was repeated 20 times and we obtained the average computation times.

\textbf{Results:} The results are shown in Table \ref{tab:computation}. We can see that the computation times increase with the increase in key size for all the algorithms and sub-protocols. Amongst the BCP algorithms, the \textbf{mDec} is computationally the most demanding while encryption is computationally the least demanding. Meanwhile, amongst the sub-protocols, the \texttt{Mult} is computationally the most demanding while \texttt{KeyProd} is compuatationally the least demanding. For the \texttt{KeyProd}, we only considered a joint public key from two users and the joint public key generation does not involve encryption of data as in \cite{peter2013}. Also, for the \texttt{TransDec}, we considered only two users. Based on the shown computation times and security requirements, we considered the key size of 1024 during the implementation of the SSXGB.

\subsection{Efficiency of SSXGB}
\begin{figure}[!t]
\begin{center}
{\includegraphics[width = 2.5in]{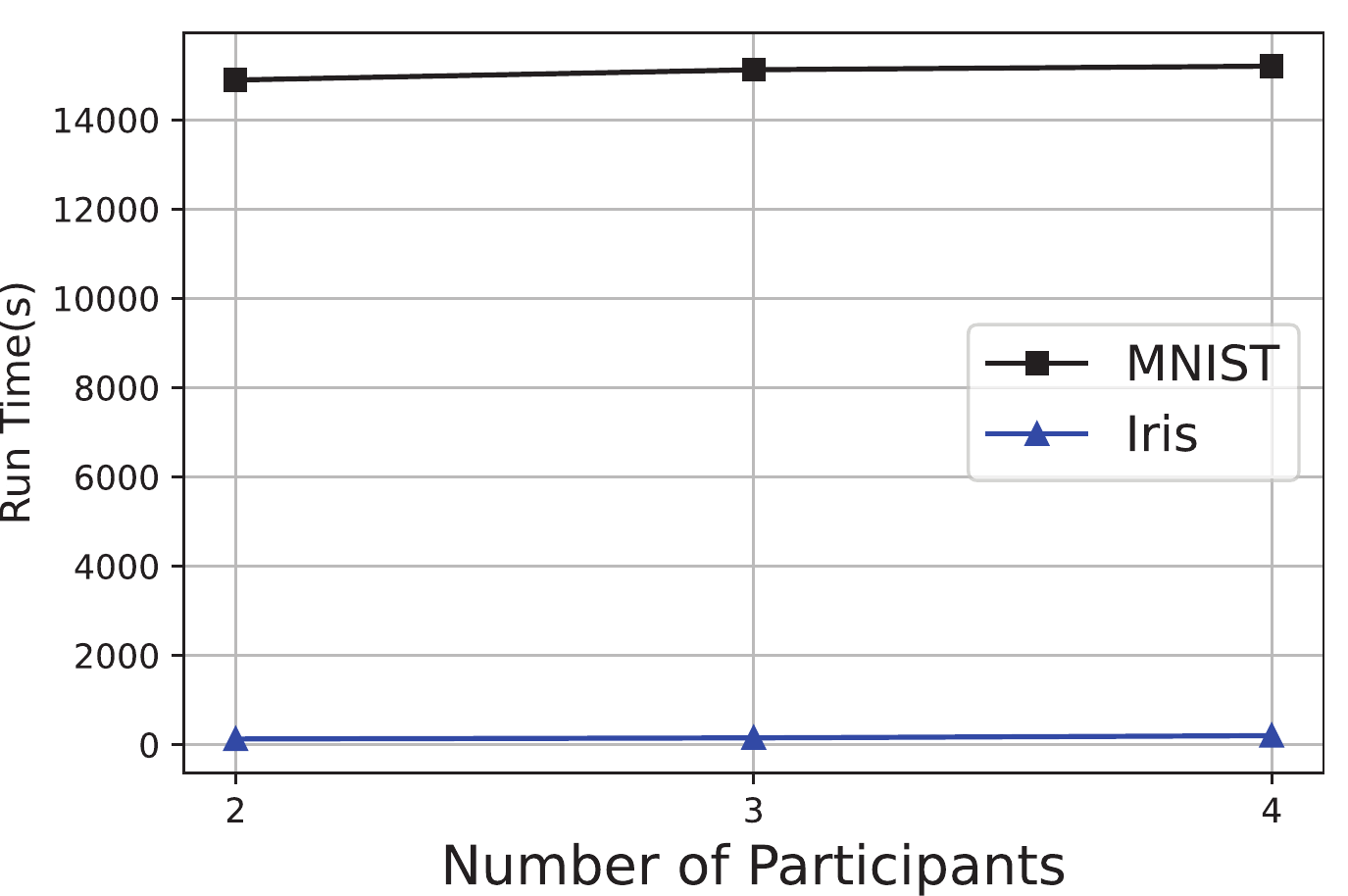}}
\caption{Running time of the SSXGB with the Iris and MNIST datasets}
\label{fig:runtime}
\end{center}
\end{figure}

Finally, we investigate the efficiency of the proposed SSXGB by examining its running time. We performed the investigation under a varying number of participants. We considered the running time for training using the two datasets. Figure \ref{fig:runtime} shows the running time of training using the MNIST and Iris datasets against the varying number of participants. Since we intended to examine the running time comparatively for the datasets and the Iris dataset has only 4 features, we limited the number of participants to only 4. And, in Figure \ref{fig:runtime}, it can be observed that there is no significant change in running time against the varying number of participants. We can attribute this to the number of features and samples remaining almost the same under the vertically partitioned datasets, hence the number of feature and sample operations remain fairly constant. However, the running time is higher for the MNIST dataset compared to the Iris dataset. This is mainly because the MNIST dataset has more features and samples as compared to the Iris dataset.

\section{Conclusion}\label{sec:conclusion}
In this work, we proposed SSXGB framework for scalable and privacy-preserving multi-party gradient tree boosting over vertically partitioned datasets with outsourced computations. We adopted BCP HE for secure computations and proposed sub-protocols based on the BCP HE for two non-linear operations of gradient tree boosting. Analysis of the framework shows that no information gets leaked to any entities under the semi-honest security model. We also implemented the secure training algorithm of the SSXGB framework. We performed experiments using two real-world datasets. The results show that the SSXGB is scalable and reasonably effective. In the future, we shall aim to minimize the communication overhead of the proposed framework.

%
%
%
%
\ifCLASSOPTIONcaptionsoff
  \newpage
\fi




\bibliographystyle{IEEEtran}
\bibliography{mybibfile.bib}

\appendices
\section{Details of the \texttt{LGT} Sub-protocol}\label{app:lgt}
\begin{figure*}[!ht]
\centering
\includegraphics[width=6.0in]{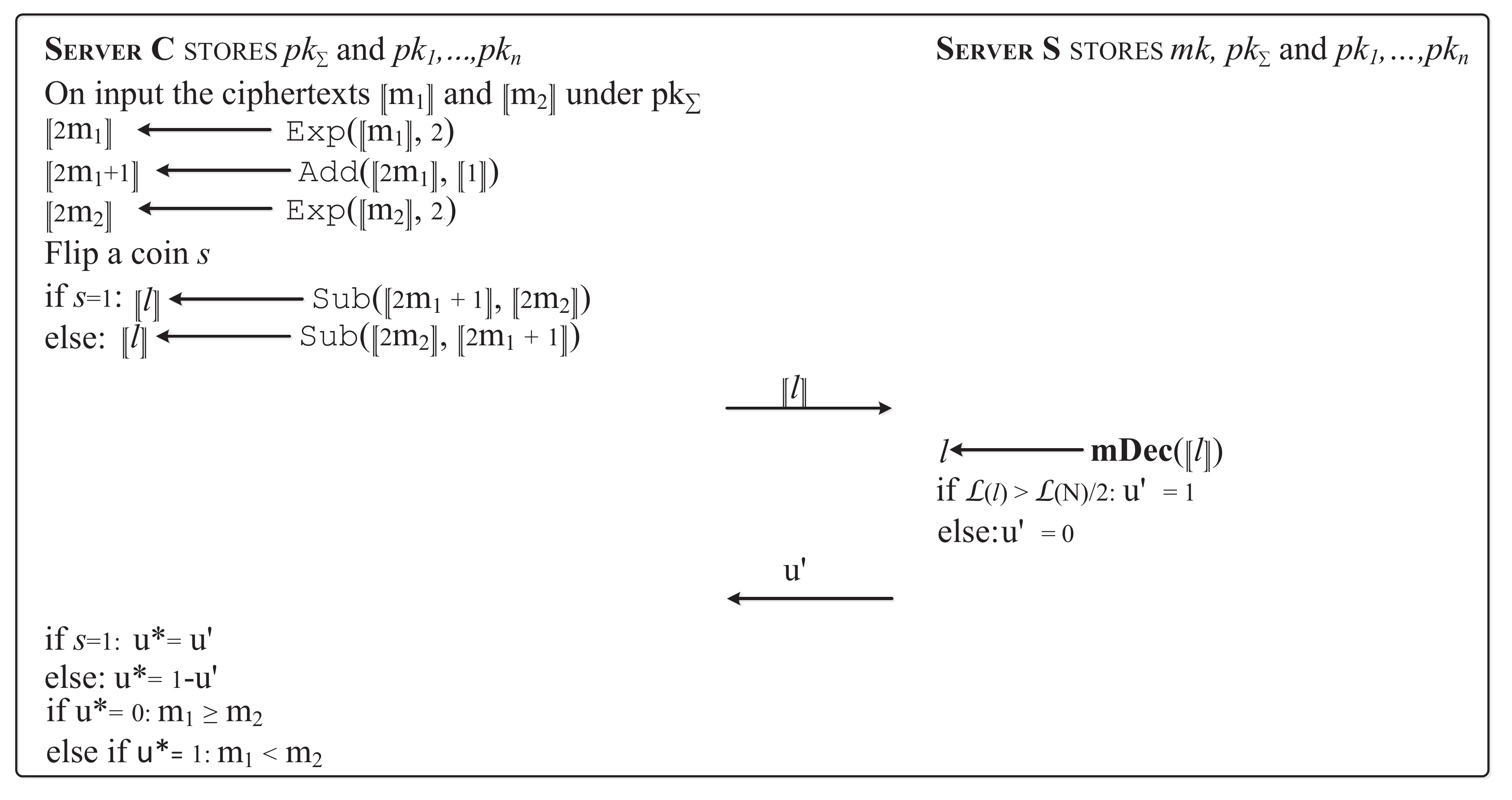}
\caption{The \texttt{LGT} sub-protocol}
\label{fig:lgt}
\end{figure*}

The \texttt{LGT} sub-protocol is an adaptation of the SLT sub-protocol in \cite{liu2016}. The only difference is that we intentionally reveal the result of the comparison to the server C. An illustration of the sub-protocol is shown in Figure \ref{fig:lgt}. Given two ciphertexts \textlbrackdbl m$_1\text{\textrbrackdbl }$ and \textlbrackdbl m$_2\text{\textrbrackdbl }$ encrypted under the joint public key $pk_{\Sigma}$. To determine if \textlbrackdbl m$_1$\textrbrackdbl$\geq$\textlbrackdbl m$_2\text{\textrbrackdbl }$ or \textlbrackdbl m$_1$\textrbrackdbl$<$\textlbrackdbl m$_2$\textrbrackdbl, the following procedures are collaboratively performed by servers C and S. 

As in \cite{liu2016}, first server C uses the \texttt{Exp} sub-protocol to multiply the ciphertexts by two as \textlbrackdbl 2m$_1\text{\textrbrackdbl }$ and \textlbrackdbl 2m$_2$\textrbrackdbl. Server C also encrypts 1 using the \textbf{Enc} algorithm as \textlbrackdbl 1\textrbrackdbl, and adds the result to \textlbrackdbl 2m$_1\text{\textrbrackdbl }$ as \textlbrackdbl 2m$_1$ + 1\textrbrackdbl.

Server C then flips a coin $s$. If $s=1$, server C uses the \texttt{Sub} sub-protocol and computes \textlbrackdbl $l\text{\textrbrackdbl }$ as (\textlbrackdbl 2m$_1$ + 1\textrbrackdbl - \textlbrackdbl 2m$_2$\textrbrackdbl). Otherwise, \textlbrackdbl $l\text{\textrbrackdbl }$ is computed as (\textlbrackdbl 2m$_2$\textrbrackdbl - \textlbrackdbl 2m$_1$+1\textrbrackdbl). Server C then sends \textlbrackdbl $l\text{\textrbrackdbl }$ to server S.

Server S uses the \textbf{mDec} algorithm to decrypt \textlbrackdbl $l\text{\textrbrackdbl }$ as $l$. Server S then computes $\mathcal{L}(l)$\footnote{$\mathcal{L}(x)$ denotes the bit length of $x$. See \cite{liu2016} for the details}. If $\mathcal{L}(l)>\mathcal{N}/2$, server S sets $u^\prime = 1$, otherwise $u^\prime = 0$. Server S then returns $u^\prime$ to server C. 

Next, server C checks for $s$, and if $s=1$, server C sets $u^* = u^\prime$. Otherwise, $u^* = 1-u^\prime$. If $u^*=0$, m$_1\geq$m$_2$, otherwise m$_1<$m$_2$.

\section{Proof of correctness for the \texttt{Div} subprotocol}\label{app:div}
Consider the division of $\text{\textlbrackdbl}m_1\text{\textrbrackdbl}$ by $\text{\textlbrackdbl}m_2\text{\textrbrackdbl}$ ($\text{\textlbrackdbl}m_1 \div m_2\text{\textrbrackdbl}$) using the \texttt{Div} subprotocol. For the purpose of this proof, we assume $\text{\textlbrackdbl}m_1\text{\textrbrackdbl}$ and $\text{\textlbrackdbl}m_2\text{\textrbrackdbl}$ are encrypted under the same public key. First, the server C, randomly selects $\tau_{_1}, \tau_{_2} \in \mathbb{Z}_N$. Using the \texttt{Exp} subprotocol, the server C generates $\text{\textlbrackdbl}\tau_{_1}m_1\text{\textrbrackdbl}$, $\text{\textlbrackdbl}\tau_{_2}m_2\text{\textrbrackdbl}$ and $\text{\textlbrackdbl}\tau_{_1}m_2\text{\textrbrackdbl}$ as follows:
\begin{equation}
\begin{split}
\text{\textlbrackdbl}\tau_{_1}m_1\text{\textrbrackdbl} &= \big{(}g^{\tau_{_1}.r_{_1}} \text{mod }N^2, h^{\tau_{_1}.r_{_1}}(1+\tau_{_1}.m_1.N) \text{mod }N^2\big{)} \\
\text{\textlbrackdbl}\tau_{_2}m_2\text{\textrbrackdbl} &= \big{(}g^{\tau_{_2}.r_{_2}} \text{mod }N^2, h^{\tau_{_2}.r_{_2}}(1+\tau_{_2}.m_2.N) \text{mod }N^2\big{)} \\
\text{\textlbrackdbl}\tau_{_1}m_2\text{\textrbrackdbl} &= \big{(}g^{\tau_{_1}.r_{_3}} \text{mod }N^2, h^{\tau_{_1}.r_{_3}}(1+\tau_{_1}.m_2.N) \text{mod }N^2\big{)}
\end{split}
\end{equation}
Using the \texttt{Add} subprotocol, the server C computes $\text{\textlbrackdbl}\tau_{_1}m_1 + \tau_{_2}m_2\text{\textrbrackdbl}$ as:
\begin{equation}
\begin{split}
\text{\textlbrackdbl}\tau_{_1}m_1 + \tau_{_2}m_2\text{\textrbrackdbl} &= \big{(}g^{(\tau_{_1}r_{_1} + \tau_{_2}r_{_2})} \text{mod }N^2, \\
& h^{(\tau_{_1}r_{_1} + \tau_{_2}r_{_2})}(1+(\tau_{_1}m_1+\tau_{_2}m_2)N) \text{mod }N^2\big{)}
\end{split}
\end{equation}
The server C then sends $\text{\textlbrackdbl}\tau_{_1}m_2\text{\textrbrackdbl}$ and $\text{\textlbrackdbl}\tau_{_1}m_1 + \tau_{_2}m_2\text{\textrbrackdbl}$ to the server S. Using the \textbf{mDec} algorithm, the server S decrypts $\text{\textlbrackdbl}\tau_{_1}m_2\text{\textrbrackdbl}$ and $\text{\textlbrackdbl}\tau_{_1}m_1 + \tau_{_2}m_2\text{\textrbrackdbl}$ as $\tau_{_1}m_2$ and $\tau_{_1}m_1 + \tau_{_2}m_2$, respectively. In plaintext, the server S then computes ($\frac{m_1}{m_2} + \frac{\tau_{_2}}{\tau_{_1}}$) as:
\begin{equation}
(\frac{m_1}{m_2} + \frac{\tau_{_2}}{\tau_{_1}}) = \frac{1}{\tau_{_1}m_2} (\tau_{_1}m_1 + \tau_{_2}m_2)
\end{equation}
Next, the server S encrypts $(\frac{m_1}{m_2} + \frac{\tau_{_2}}{\tau_{_1}})$ to $\text{\textlbrackdbl}\frac{m_1}{m_2} + \frac{\tau_{_2}}{\tau_{_1}}\text{\textrbrackdbl}$ as:
\begin{equation}
  \text{\textlbrackdbl}\frac{m_1}{m_2} + \frac{\tau_{_2}}{\tau_{_1}}\text{\textrbrackdbl} = \big{(}g^{r} \text{mod }N^2, h^{r}(1+(\frac{m_1}{m_2} + \frac{\tau_{_2}}{\tau_{_1}}).N) \text{mod }N^2\big{)}
\end{equation}
and sends the result to server C.

After receiving $\text{\textlbrackdbl}\frac{m_1}{m_2} + \frac{\tau_{_2}}{\tau_{_1}}\text{\textrbrackdbl}$, the server C computes $\frac{\tau_{_2}}{\tau_{_1}}$ in plaintext and encrypts it as $\text{\textlbrackdbl}\frac{\tau_{_2}}{\tau_{_1}}\text{\textrbrackdbl}$. The server C extracts $\text{\textlbrackdbl}\frac{m_1}{m_2}\text{\textrbrackdbl}$ using the \texttt{Sub} subprotocol as:
\begin{equation}
  \text{\textlbrackdbl}\frac{m_1}{m_2}\text{\textrbrackdbl} = \text{\texttt{Sub}}(\text{\textlbrackdbl}\frac{m_1}{m_2} + \frac{\tau_{_2}}{\tau_{_1}}\text{\textrbrackdbl}, \text{\textlbrackdbl}\frac{\tau_{_2}}{\tau_{_1}}\text{\textrbrackdbl}).
\end{equation}

%

\end{document}